\documentclass[12pt,preprint]{aastex}

\shorttitle{OPTICAL VARIABILITY OF L DWARFS}
\shortauthors{Maiti et al.}
\newcommand{\beqna}{\begin{eqnarray}}
\newcommand{\eeqna}{\end{eqnarray}}
\begin{document}

\title{OBSERVATION OF R-BAND VARIABILITY OF L DWARFS}
\author{M. Maiti\footnote{e-mail: mith@iiap.ernet.in}, S. Sengupta,
P. S. Parihar \and G. C. Anupama}
\affil{Indian Institute of Astrophysics, Koramangala, Bangalore-560 034}

\begin{abstract}
We report, for the first time, photometric variability of L dwarfs in $R$ band.
Out of three L1 dwarfs (2MASS 1300+19, 2MASS 1439+19, and 2MASS
1658+70) observed, we have detected R band variability in 2MASS 1300+19 and
2MASS 1439+19. The objects exhibit variability of amplitude ranging from 0.01
mag to 0.02 mag.  Object 2MASS 1658+70, turns out to be non-variable  
in both $R$ and $I$ band. However, more observations are needed to infer its
variability.  No periodic behaviour in the variability is found from the two
L1 dwarfs that are variable.  All the three L1 dwarfs have either negligible 
or no $H_{\alpha}$ activity. In the absence of any direct evidence for the
presence of sufficiently strong magnetic field, the detection of polarization
at the optical favors the presence of dust in the atmosphere of L dwarfs.
We suggest that the observed $R$ band photometric variability is most likely
due to atmospheric dust activity. 

\end{abstract}


\section{INTRODUCTION}
L dwarfs are ultra-cool objects with effective temperature
ranging between 2200 K and 1400K. They are characterized by the presence of
condensates in their atmosphere. Due to incomplete gravitational settling,
dust in the atmosphere of L dwarfs could be detectable in the optical. Dust
cloud can play a potential role in making the object variable. Time-resolved
photometric variability
of a large number of L dwarfs has been reported by Bailer-Jones \& Mundt 
(2001a,b), Martin, Osorio, \& Lehto (2001), Gelino et al. (2002), Clarke,
Oppenheimer \& Tinney (2002a),
Osorio, Caballero, Bejar \& Rebolo (2003), Bailer-Jones \& Lamm (2003).
However, all these observations were made in the $I$ and $J$ bands. 
Clarke, Tinney \& Covey (2002b) reported variability from ultra-cool dwarfs 
by using a non-standard filter with the effective wavelength similar to that
of $I$ band.  Enoch, Brown \& Burgasser (2003) reported evidence of 
variability in $K_s$ band from a few
L and T dwarfs.  These investigations provide much insight on the atmospheric
activities, especially the presence of dust clouds.

Sengupta \& Krishan (2001) argued that the presence of dust
could give rise to a detectable amount of linear polarization in the optical
from L dwarfs. This was observationally verified by Menard, Delfosse \& Monin
(2002) who detected non-zero linear polarization at red (0.768 $\mu$~m) from
a few L dwarfs. In the absence of any direct evidence of sufficiently strong
magnetic field, observation of linear polarization strongly favours the 
presence of dust in the atmosphere of L dwarfs and single dust scattering
model could explain the observed polarization (Sengupta 2003).
It should also be mentioned here that rigorous theoretical analysis 
(see Burrows et al. 2001 and references therein; Tsuji et al. 2004) of the
continuum spectra implies the presence of dust in the visible region of L
dwarfs.

Detection of non-zero polarization at the optical band that may arise from
dust scattering urges the necessity of investigating $R$ band variability by
dust activity in L dwarfs. In the present letter we, for the first time,
report differential photometric variability in $R$ band from a few L dwarfs.
The results could be a good complement to the polarization observations as
dust clouds play a crucial role in both cases. A detailed theoretical
investigation on polarization by single dust scattering of L dwarfs with fixed 
rotational velocity (Sengupta \& Kwok 2004) shows that the degree of 
linear polarization peaks at L1 spectral 
type. This motivates us to concentrate on the observation of L1 dwarfs. We
have detected photometric variability from two L1 dwarfs. In the next
section we describe the observation and data reduction procedure followed.
The results are presented and discussed in section~3 followed by conclusions.

\section{OBSERVATION AND DATA REDUCTION}
The photometric observations of selected L dwarfs were carried out
during 2004 January -- June
using the 2-m Himalayan Chandra Telescope (HCT) of the Indian Astronomical 
Observatory (IAO) at Hanle, India, using the Himalaya Faint Object
Spectrograph Camera (HFOSC), equipped with a SITe $2\times 4$~K pixel
CCD. The central $2\times 2$~K region used for imaging corresponds to a field 
of view of $10\,{\rm{arcmin}}\times 10\,{\rm{arcmin}}$ at 0.296~arcsec 
pixel$^{-1}$. More details on the telescope and the instrument
may be obtained from {\it{http://www.iiap.res.in/$\sim$iao}}.

We selected the targets from published spectroscopically determined
L dwarfs, specifically selecting those objects that have negligible or no
$H_{\alpha}$ effective line-width. This avoids the contribution of $H_{\alpha}$
line variability to the atmospheric variability in the $R$ band and hence any
variability observed could be attributed more convincingly, to the presence of
dust clouds.  Table 1 presents the name, position, and the $R$ and $I$ 
magnitudes of those objects.  

Several exposures with times of 10~minutes and 5~minutes were 
obtained in the $R$ and $I$ bands respectively. The central wavelengths of 
the R band filter and the I band filter used are 0.6 and 0.805 micron
respectively.  To minimize the effect of 
improper flat fielding and any systematic error spatially associated with the 
chip, we tried to confine the L dwarf to a particular CCD pixel in all
the frames. The observations were carried out during dark moon period,
and uninterrupted observations of a single object were obtained over 3 to 7 
hours during different nights. The full width at half maximum (FWHM) of the
stellar profile was found to be about 1.5 to 1.8  arc-seconds.
The observing log is given in the Table \ref{tab2}.
The basic image processing such as bias subtraction and flat fielding were 
done in the standard manner using the various tasks available within IRAF. 
Atmospheric extinction and transformation coefficients were obtained from
observation of photometric standard stars (Landolt 1992). The $I$ frames were
affected by CCD fringing due to night sky emission lines. Fringe correction
was applied to all the $I$ frames using a master fringe frame created by
observing several blank night sky fields.

The stellar magnitudes at varying apertures were then obtained using the
IRAF task \textit{phot}. Since accurate 
sky background estimation is very crucial to faint object photometry, the
sky value was iteratively estimated by examining the growth curves of 
isolated stars so that the growth curves were neither monotonically 
decreasing (underestimated sky) nor increasing (overestimated sky).
Furthermore, the magnitudes of L dwarfs and faint stars
were first determined at the aperture having  highest $S/N$ and then
aperture correction was made to the aperture size $4\times FWHM$ using
the correction term obtained from bright isolated stars.
The standard $R$ and $I$ magnitudes of L dwarfs and the field stars were 
determined using system transformation coefficients.

Differential photometry was performed following the ensemble photometry 
technique (Gilliland \& Brown 1988; Everett \& Howell 2001). From the average
flux of the few fairly bright stars we determined the ensemble reference
magnitude, iteratively rejecting stars that were found to have either a 
systematic variation or large errors. 
The differential magnitude of L dwarfs with respect to the ensemble magnitude
was then computed using the relation
\begin{eqnarray}
         \Delta m_{i,b} & = & \overline m - m_{i,b},
\end{eqnarray}
where $\overline m$ is the ensemble magnitude and $m_{i,b}$ is the magnitude of
L dwarfs.

 While analyzing the differential photometric data of L dwarfs, a linear
variation with respect to airmass was noticed.
This trend appears to be an effect of the second order  extinction
coefficient. The spectral type of the observed ultra-cool dwarfs were L1
whereas our ensemble references were found to be near spectral type G. 
Therefore, within our  observing band they would
have very different effective wavelengths. In order to check  how severe the 
second order extinction effect  would be,  we first  determined the effective  
wavelengths  of the L dwarfs and G type reference stars using equation:
\begin{eqnarray}
     \lambda_{eff} & = & \frac{ \int \lambda~F_{\lambda}~S_{\lambda}~d\lambda} 
                          {\int F_{\lambda}~S_{\lambda}~d\lambda}
\end{eqnarray} 
\noindent
 where $F_{\lambda}$ is spectral energy distribution and $S_{\lambda}$ is 
system response. The L dwarf and reference star spectra were
retrieved from digital spectral libraries (Martin, Delfosse \& Basri 1999;
Le Borgne et al. 2003). The computed effective wavelengths for L 
dwarf and G type star are 6308~\AA\  and 7148~\AA\  respectively. 
The average spectroscopic extinction at IAO
are $0.081~mag$ and $0.04~mag$  at these two wavelengths.   
The difference  of these  two extinction values is nothing but 
the second order extinction correction at the unit airmass i.e.
$k^{''}_{R}\Delta(R-I)$. The observed color difference $(R-I)$ of
L dwarf and G type reference star was found to be $2.1~mag$ and
that predicts $k^{''}_{R}\approx -0.02~mag$. 
The second order extinction coefficient was independently estimated using the
observations of the photometric standards and found to be
$k^{''}_R \approx -0.02$, similar to the above estimation . The second order 
extinction correction was made to each observed L dwarf data using the relation 
\begin{eqnarray}
         (\Delta m_{i,b})_o & = & \Delta m_{i,b} - k^{''}_R  \Delta (R-I) X_i
\end{eqnarray}
\noindent where $X_i$ is airmass of the frames.

\section{RESULTS AND DISCUSSIONS}

The light curves of the three L1 dwarfs observed are presented in figure 1
and in figure 2. Figure~1 shows the light curves of 2MASS 1300+19 and
2MASS 1439+19 in $R$ band, while figure 2 shows the light curves of
2MASS 1658+70 in both $R$ and $I$ bands. The light curves in $R$ band for
all the three objects indicate a possible variability. 

In order to verify any variability, we employ the procedure given by
Martin et al.~(2001). In figure~3a-e , we show the standard deviation, 
$\sigma_R$ with respect to the standard R magnitudes of all the  field
stars for different fields at different nights along with that for the three
targets. 
Figure 3a-d show that although the programme stars 2MASS 1300+19 and 2MASS
1439+19 lie at the variable side of the
$\sigma_R$ vs the R magnitude diagram, no overwhelming evidence for variability
in any of the two objects is found. On the other hand Figure 3e shows that
the programme object 2MASS 1658+70 lies at the non-variable side of the
diagram for 19 May implying no variability in R band. A similar inference
can be made from the other observing nights for R band as well as for
the I-band observation of the same object.

From figure 3a-e, we notice that for the same field there are very less 
variation of systematic error at different nights. Also, for different fields
the relation between $\sigma_R$ with standard R magnitudes and its 1$\sigma$ 
scatter do not change significantly. In order to have
objects well distributed across the entire magnitude range  we combine the
different fields. This should provide a statistically
more reliable result for the $\sigma_R$ vs R magnitude relationship.
The result with combined fields is presented in figure~3f.
For this case, the relation between $\sigma_R$ vs standard R magnitudes can be
written as 
\begin{eqnarray}\label{rel}
\sigma_R = 0.712139 - 0.0854801~R + 0.00256972~R^2
\end{eqnarray}
with a scatter of
0.004 ($1\sigma$). However, the results do not differ from that obtained by
using the individual fields. The objects
2MASS 1300+19 and 2MASS 1439+19 are found to be nearly $2\sigma$ and
$1\sigma$ away from the mean relation respectively.

On the other hand, if we determine the systematic error from the relation 
between $\sigma_R$ vs  standard R magnitude and consider its  large $1\sigma$
error as well, the systematic error becomes about double the photometric
error which is calculated using  ensemble references and program star 
photometric errors. Even if we add the scatter due to color effect, such a
large error is not expected. However, if we  discard the objects that are
fairly out of fit in figure \ref{fig3}(f) as well as having systematic trend
in their light curve and consider rest of the  field objects instead, the
scatter of distribution of error reduces to 0.002 ($1\sigma$). Consequently,
the objects 2MASS 1300+19 and 2MASS 1439+19 are found to be situated at about
5$\sigma$ and 3$\sigma$ away from the mean relation respectively.  Therefore,
our analysis implies variability of 2MASS 1300+19 and 2MASS 1439+19 in R band.

The statistical significance of the observed variability is also checked by
computing the $\chi^2$ by  using the formula:
\begin{eqnarray}
         \chi^2 = \sum_{i = 1}^n \left( \frac{\overline {\Delta m_b } -
\Delta m_{i,b}}{\sigma_{i,b}} \right)^2
\end{eqnarray}
\noindent
\noindent where \textit{n} is the number of observed data points and
 $\sigma_{i,b}$ is the error associated with the L dwarfs at their respective
magnitudes, as obtained from the standard deviation versus standard magnitudes
relation given by equation (\ref{rel}).
However, it is worth mentioning that the ratio between the variance of L
dwarf data and the variance read from the fitted curve may not have a
$\chi^2$ distribution and hence it should be considered as an assumption.
Table~\ref{tab2} gives the 
results for each object for individual nights of observation, the number of good
frames taken for final analysis, number of references taken to make the mean
standard, the standard deviation of the points from the mean level
($\sigma_{rms}$), the average $\sigma _{rms}$ used for $\chi^{2}$ test  and
the probability that the L dwarf is variable ($p$).
The L dwarfs 2MASS 1439+19 and 2MASS 1300+19 indicate variability with about
99\% probability for all the observed nights. Note that $\sigma_{i,b}$ are 
calculated from the $\sigma_R$ vs R magnitude relation by considering all the
field stars including those that show systematic trend in their light curve.

A period analysis program based on the widely used Scargle formalism
(Scargle 1982) was used to search for any periodicity in our time series
photometric data. However, we have not obtained any significant periodicity in 
the variations from either L dwarfs.

The third object, 2MASS 1658+70, does not show any 
variability both in $R$ and $I$ bands.  However an inspection of the light
curves (see figure~\ref{fig2}) of the 20th of  May  and the 16th of  June
although look like scatter plot, there is a systematic trend in the light
curves of the 19th and the 21st May 2004 suggesting a possibility of variation.
Further, Gelino et al (2002) reported this object to be variable in $I$ band.
It is therefore possible that the variability in this object is transient due
to the dust-active variation.  More observations with improved temporal
coverage are required to establish the variable nature of 2MASS 1658+70.

\section{CONCLUSIONS}
$R$ band differential photometry of three L1 dwarfs 2MASS 1300+19, 2MASS
1439+19, and 2MASS 1658+70 are presented here. The first two objects show
variability. However, no  periodicity in the 
variation is obtained in any of these objects. The light curves indicate 
transient activity of short time scale. Since rotationally related
variability should show rather smooth light curves, any co-relation
with the rotation of the objects is unlikely. The third object,
2MASS 1658+70, which was also observed in the $I$ band does not show any 
variability either in the $R$ or in the $I$ band although it was reported to 
be a variable in the $I$ band by Gelino et al.~(2002). The light
curves of this object presented here, however, provide an indication of 
possible flux variation. More observations of this object are required before
its variable nature can be established.

The observation of linear polarization in the red by Menard, Delfosse \&
Monin (2002) favours the presence of dust cloud in the atmosphere of L dwarfs.
The synthetic spectra of L dwarfs also favours the presence of dust in the
atmosphere.  We propose that the photometric variability in the $R$ band
reported here arises due to the dust activity in the atmosphere.

On the other hand, if the variability observed is due to the dust activity, 
then we predict non-zero polarization from L1 dwarfs at R band
by dust scattering.  However, photospheric variability, if caused by dust
cloud, needs sufficiently optically thick dust layer. In an optically thick
medium, polarization would arise by multiple scattering of photons. It is
known that multiple scattering reduces the degree of polarization as
compared to that by single scattering mechanism (Sengupta 2001, 2003). Hence,
a variable L dwarf should show less amount of polarization as compared to that
of non-variable or weakly variable objects which might have optically thin
dust layer. 

 Further observations of L dwarfs of different spectral types 
and in both $R$ and $I$ bands could tell if there is any co-relation in the
variability at $R$ and $I$ bands with the spectral type, which in turn could 
provide significant insight on the distribution of dust in the atmosphere of
L dwarfs. 

\acknowledgments
We are thankful to the referee for many useful suggestions, comments and
constructive criticism.
Thanks are due to A. V. Raveendran, A. Saha, and T. P. Prabhu for several
discussions.

The Observations reported in this letter were obtained using the 2-m
Himalayan Chandra Telescope  at Mt. Saraswati, Hanle, Indian Astronomical
Observatory, the high altitude station of the Indian Institute of
Astrophysics, Bangalore.
We  thank the staff at IAO and at the remote control station  at CREST,
Hosakote for assistance during the observations.

\begin{figure}[hb]
\includegraphics[scale = .9, trim = 0 150 00 00,clip]{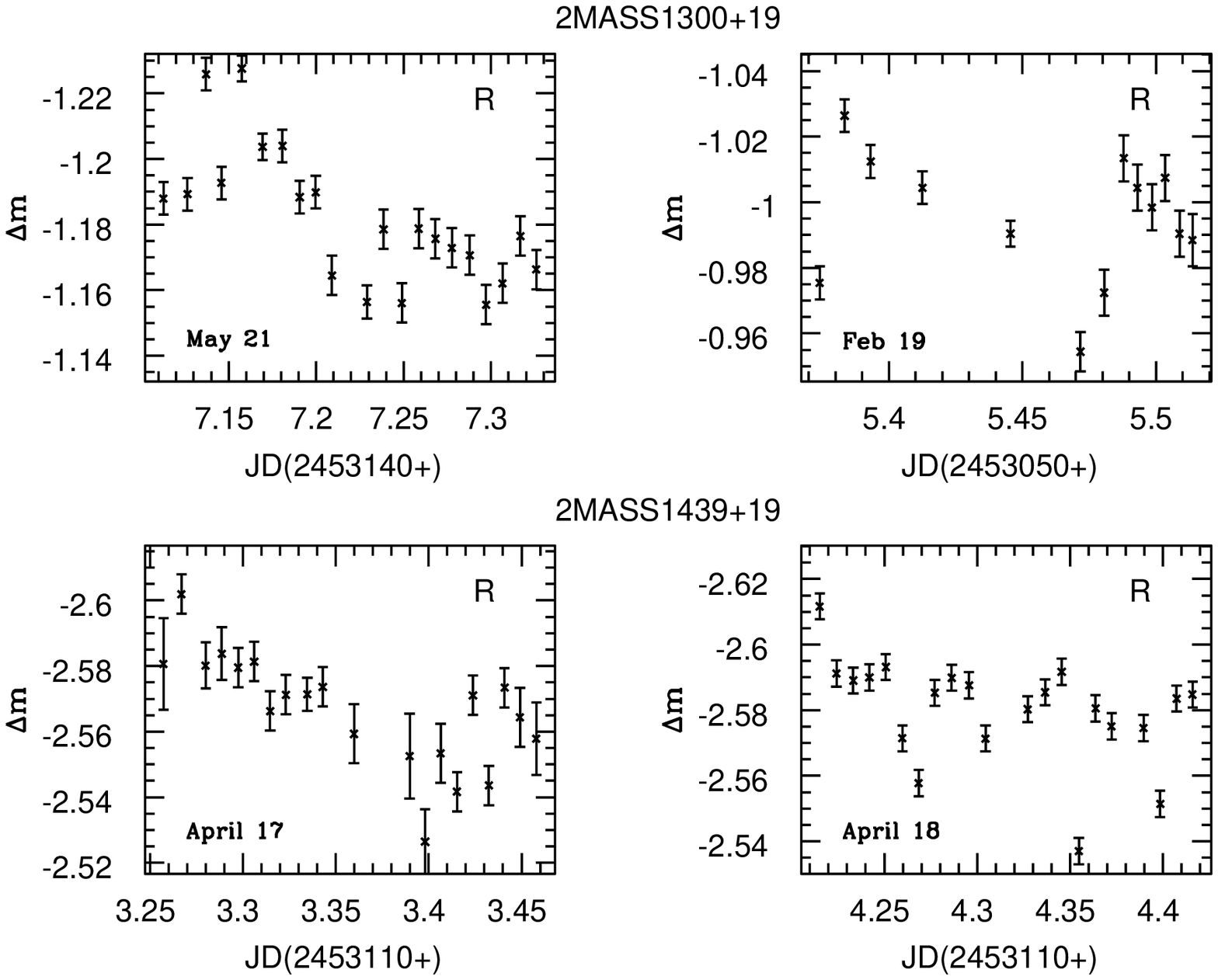}
\caption{The $R$ band light curves of L1 dwarfs 2MASS1300+19 (top) and
2MASS1439+19 (bottom) obtained on different nights.\label{fig1}}
\end{figure}

\clearpage

\begin{figure}[hb]
\includegraphics[scale = .9, trim = 0 150 00 00,clip]{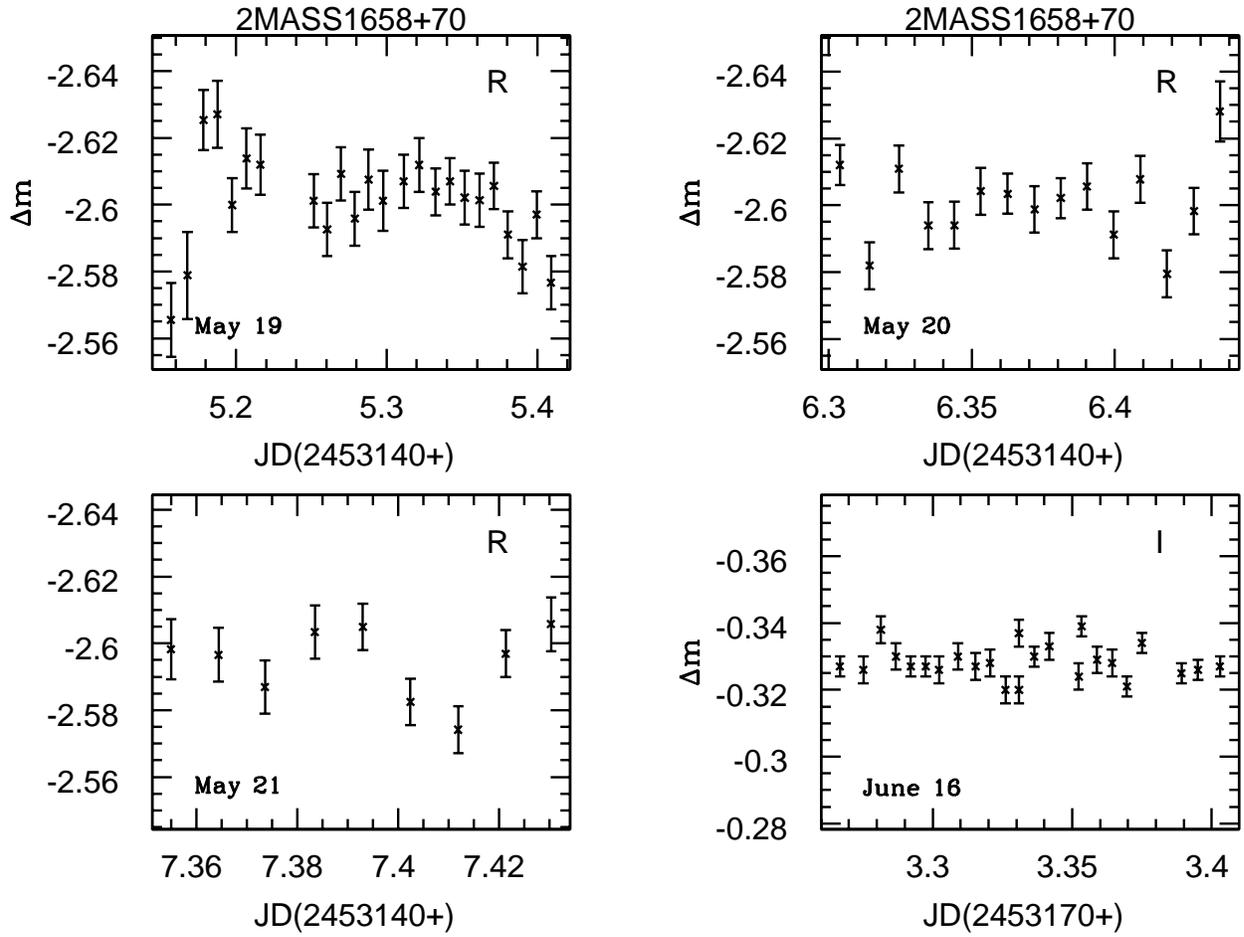}
\caption{The $R$ and $I$ band light curves of the L1 dwarf 2MASS1658+70 
obtained on different nights. \label{fig2}}
\end{figure}

\clearpage
\begin{figure}
\includegraphics[scale = .9, trim = 0 0 00 00,clip]{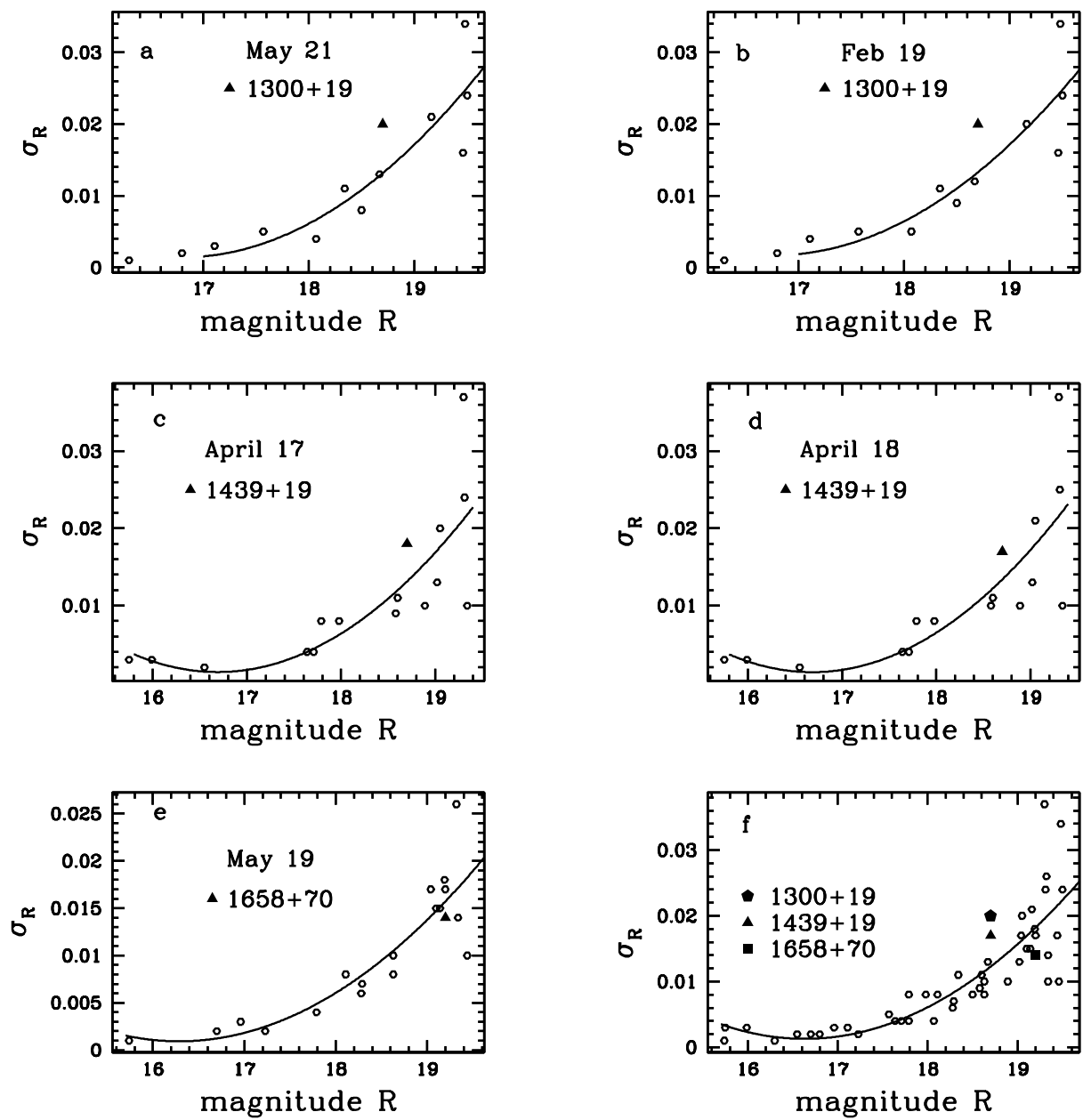}
\caption{Standard deviation ($\sigma_R$) versus $R$ band magnitudes for
the field stars for three different fields for individual nights.
A second-order polynomial fit to the $\sigma_R$   
is shown as a continuous curve. The  $\sigma_{rms}$ on each night
for all the three L1 dwarfs are also shown as filled symbols. The result with
combined data of May21, April 17 and May 19 is given in plot f.
\label{fig3}}
\end{figure}
\clearpage

\clearpage

\begin{table}
\begin{center}
\caption{BASIC DATA OF VARIABLE L1 DWARFS. \label{tab1}}
\vskip5mm
\begin{tabular}{ccccccc}
\tableline\tableline
Name &  RA(2000) & DEC(2000) & $R~mag$ & $I~mag$ & $H_{\alpha}$ Emission \\
\tableline
2MASS 1300+19  &  13 00 42.5 & +19 12 35 & 18.7 & 16.0  & ... \\                 
2MASS 1439+19  &  14 39 28.4 & +19 29 15 & 18.7 & 16.0  & $< 0.03$ \\              
2MASS 1658+70  &  16 58 03.7 & +70 27 01 & 19.2 & 16.6   & ... \\
\tableline
\end{tabular}
\end{center}
\end{table}
\clearpage
\begin{table}[h]
\begin{center}
\caption{Observing log and results \label{tab2} }
\begin{tabular}{cccccccc}
\tableline\tableline
Name & Dates & Frames  & Ref. &  Filter
& $\sigma_{rms}$\tablenotemark{a} of & $\sigma_{rms}$\tablenotemark{b} used   &
p(\%) \\
&  & (Used/Total)  & stars  &  & L dwarfs &  in $\chi^2$  & \\
\tableline
2MASS1300+19 & 19/02/04 &  13/15 &  5 & R &  0.019  &   0.012 & $>$99.0\\
2MASS1300+19 & 21/05/04 &  21/24 &  5 & R &  0.020  &   0.012 &  $>$99.0\\
\hline
2MASS1439+19 & 17/04/04 &  20/24 &  5 & R &  0.017  &   0.012 &  $>$99.0\\
2MASS1439+19 & 18/04/04 &  21/24 &  5 & R &  0.016  &   0.012 &  $>$98.5\\
\hline
2MASS1658+70 & 19/05/04 &  24/25 &  5 & R &  0.014  &   0.018 &  10.1\\
2MASS1658+70 & 20/05/04 &  15/15 &  5 & R &  0.012  &   0.018 &  4.5\\
2MASS1658+70 & 21/05/04 &  9/9   &  5 & R &  0.011  &   0.018 &  6.1\\
2MASS1658+70 & 16/06/04 &  24/24 &  3 & I &  0.005  &   0.007 &  3.1\\
\tableline
\end{tabular}
\tablenotetext{a}{$\sigma_{rms}$ about the mean differential magnitude of the
L dwarf.}
\tablenotetext{b}{$\sigma_{rms}$ used  in $\chi^2$ test, obtained from standard
deviation ($\sigma_R$) vs. standard  R magnitudes relation.}
\end{center}
\end{table}

\end{document}